\documentclass[paper,12pt]{JHEP3} 

\title{Deep inelastic scattering from gauge string duality in the soft wall model}

\author{C.  A.  Ballon Bayona, Henrique Boschi-Filho and Nelson R. F. Braga \\
Instituto de F\'{\i}sica,
Universidade Federal do Rio de Janeiro, Caixa Postal 68528, RJ
21941-972 -- Brazil\\
E-mails: \email{ballon@if.ufrj.br}, \email{boschi@if.ufrj.br}, \email{braga@if.ufrj.br}}

\preprint{arXiv:0711.0221 [hep-th] }

\abstract{Deep inelastic structure functions have been calculated by Polchinski and Strassler in  gauge/string duality introducing a hard infrared (IR) cut off in AdS space.
Here we investigate this problem using a soft IR cut off that leads to linear Regge trajectories for mesons.
We calculate the structure functions for scalar particles in the large $x$ regime where supergravity approximation holds
and the small $x$ regime where massive string states contribute. We also propose a hybrid model to calculate 
structure functions for fermions in the supergravity approximation.
In the deep inelastic limit our results are in agreement with those obtained using a hard cut off.}


\begin{document}

\section{ Introduction }

Gauge/string dualities inspired in AdS/CFT have provided recently many important results 
concerning the description of strong interactions. 
The AdS/CFT correspondence \cite{Maldacena:1997re,Gubser:1998bc,Witten:1998qj}
is an exact duality between a string theory in
ten dimensions and a superconformal gauge theory in a lower dimensional space.
In particular, it relates string theory in $AdS_5 \times S^5 $ space to  ${\cal N} = 4$ 
Yang Mills $SU(N)$ theory with large $N$ in four dimensions. 
Other exact gauge string dualities\cite{Klebanov:2000hb,Maldacena:2000yy}
relate non conformal ${\cal N }= 1$ gauge theories to string theory in less symmetric geometries. 

Approximate dualities have been proposed such that the gauge theories have some properties similar to QCD, the so called AdS/QCD approach. Polchinski and Strassler\cite{Polchinski:2001tt,Polchinski:2002jw} introduced an infrared cut off in the gauge theory by considering an AdS slice with  a size related to this cut off, 
the now called hard wall model. Using this approach they found the correct high energy 
scaling of hadronic  amplitudes for fixed angle scattering\cite{Polchinski:2001tt}. 
This scaling was observed experimentally and also reproduced by QCD a long time ago\cite{QCD1,BRO} but was in contrast to string theory predictions in flat space. 
This scaling was also analyzed in the  gauge/string duality approach in 
\cite{Giddings:2002cd,BoschiFilho:2002zs,Brower:2002er,Andreev:2002aw}.

The introduction of an infrared cut off in the AdS space leads to a discrete spectrum
for normalizable fields. It is natural to associate these bulk modes with
boundary masses, so the hard wall model is useful to estimate hadronic masses
\cite{Boschi-Filho:2002vd,deTeramond:2005su,Erlich:2005qh,DaRold:2005zs}. On the other side, the observed spectrum of hadrons is such that the states exhibit approximate linear relations between mass squared and spin (or radial number), the so called Regge trajectories. The hard wall model does not predict this linear behavior but rather asymptotically quadratic trajectories. This motivated a different AdS/QCD approach consisting of a background involving AdS space and a dilaton field. This field acts effectively as a smooth  infrared cut off and leads to linear Regge trajectories for mesons\cite{Karch:2006pv} and glueballs\cite{Colangelo:2007pt}. This is the so called soft wall model. For fermions the soft wall model does not lead to a discrete spectrum since the dilaton introduced in the action factors out in the equations of motion\cite{Forkel:2007cm}.
 
The hard and soft wall models describe holographically confining gauge theories.
At finite temperature, both models have a gravity phase transition that corresponds to a confinement/deconfinement transition. However the transition occurs at different temperatures for these models\cite{Herzog:2006ra} (see also\cite{BallonBayona:2007vp}). 

A very important process that provides information on the hadronic structure is the deep inelastic scattering (DIS) \cite{Manohar:1992tz}. 
A detailed description of this process using gauge/gravity duality was formulated in 
\cite{Polchinski:2002jw} in the hard wall model (see also \cite{Hatta:2007he}). 
The structure functions were obtained from string theory in different regimes
of the Bjorken parameter $x$ for the case of large 't Hooft parameter $gN$.
Gauge string duality has also been used to calculate hadronic form factors 
\cite{Grigoryan:2007vg,Grigoryan:2007my,Brodsky:2007hb}. Also, very recently, the problem of deep inelastic scattering in ${\cal N} = 4\, $ SYM plasma at strong coupling, in the context of gauge/string duality, was discussed in \cite{Hatta:2007cs}.
 
Since the soft and hard wall models predict different behaviors for some physical quantities, one could also expect different structure functions for deep inelastic scattering. The proposal of this article is to calculate these structure functions  
for the soft wall model in the case of large 't Hooft parameter $gN$. 
This is done for the scalar fields which are normalizable in this model. 
We consider two different regimes: $ 1 > x > >  (gN)^{-1/2}\,$ corresponding to  massless string excitations (supergravity approximation) and 
$  \exp \,( - \sqrt{gN }) <<  x < <  (gN)^{-1/2}\,$ corresponding to massive string excitations. 

We also discuss the fermionic case where the soft wall dilaton background is not enough to normalize the fields. We propose a different model combining hard and soft cut offs to calculate fermionic structure functions.  

In section {\bf 2} we will present the gauge string duality approach to deep inelastic scattering. 
In section {\bf 3} we will calculate the structure functions for scalar particles in the soft wall dilaton background within the supergravity approximation. In section {\bf 4} we  study the contribution of the massive string excitations to the hadronic structure functions in this model. In section {\bf 5} we present our conclusions. In the appendix we present our hybrid model for fermions.

%%%%%%%%%%%%%%%%%%%%%%%%%%%%%%%%%%%%%%%%%%%%%%%%%%%%%%%%%%%%%%%%%%%%%%%%%

\section{ Deep Inelastic Scattering and gauge string duality }

Deep inelastic scattering consists of the scattering of a lepton from a hadron.
The lepton produces a virtual photon of momentum $q^\mu$ which interacts with the hadron of momentum $P^\mu $. The final hadronic state, represented by $X$ with momentum $P_X^\mu$, is not observed (see Fig.~1). The experiment detects the final lepton, determining the momentum transfer $q^\mu$, but not the final hadronic state $X$. Then the corresponding inclusive cross section involves a sum over all possible $X$.     
We can parametrize the process using as dynamical variables the photon virtuality $q^2$ and the Bjorken parameter $x \equiv  -q^2 /2P\cdot q \,$. Deep inelastic scattering corresponds to the limit $q^2\to\infty$, with $x$ fixed.

%%%%%%%%%%%%%%%%%%%%%%%%%%%%%%%%%%%%%%%%%%%%%%%%%%%%%%%%%%
%%%%%%%%%%%%%%%%%%%%%%%%%%%%%%%%%%%%%%%%%%%%%%%%%%%%%%%%%%
%%% figura do espalhamento profundamente inelastico   %%%%
%%%%%%%%%%%%%%%%%%%%%%%%%%%%%%%%%%%%%%%%%%%%%%%%%%%%%%%%%%
%%%%%%%%%%%%%%%%%%%%%%%%%%%%%%%%%%%%%%%%%%%%%%%%%%%%%%%%%%
\begin{figure}\begin{center}
\setlength{\unitlength}{0.1in}
\vskip 3.cm
\begin{picture}(0,0)(15,0)
\rm
%%%%%%%%%%%%%%%%%%% Lepton %%%%%%%%%%%%%%%%%%%%%%
\thicklines
\put(1,14.5){$\ell$}
\put(3,15){\line(2,-1){7}}
\put(3,15){\vector(2,-1){4}}
\put(18,14.5){$\ell$}
\put(17,15){\line(-2,-1){7}}
\put(10,11.5){\vector(2,1){4.2}}
%%%%%%%%%%%%%%%%%%%  Foton  %%%%%%%%%%%%%%%%%%%%%%%
\put(9.5,8){$q$}
\bezier{300}(10,11.5)(10.2,10.7)(11,10.5)
\bezier{300}(11,10.5)(11.8,10.3)(12,9.5)
\bezier{300}(12,9.5)(12.2,8.7)(13,8.5)
\bezier{300}(13,8.5)(13.8,8.3)(14,7.5)
%%%%%%%%%%%%%%%%%%   Proton  %%%%%%%%%%%%%%%%%%%%%%%
\put(0,-2){$P$}
\put(3,0){\line(2,1){10.5}}
\put(3,0){\vector(2,1){6}}
%%%%%%%%%%%%%%%%%% Interaction %%%%%%%%%%%%%%%%%%%%%
\put(16,6){\circle{5}}
%%%%%%%%%%%%%%%%%%   Hadrons %%%%%%%%%%%%%%%%%%%%%%%
\put(27,-2){$X$}
\put(18.5,5.5){\line(3,-1){8}}
\put(18.3,5){\line(2,-1){8}}
\put(18,4.5){\line(3,-2){7.5}}
\put(17.5,3.8){\line(1,-1){6}}
%%%%%%%%%%%%%%%%%%%%%%%%%%%%%%%%%%%%%%%%%%%%%%%%%%%%%
\end{picture}
\vskip 1.cm
\parbox{4.1 in}{\caption{} Illustrative diagram for a deep inelastic scattering. A lepton $\ell$ exchanges a virtual photon with a hadron of momentum $P$.}
\end{center}
\end{figure}
\vskip .5cm
%%%%%%%%%%%%%%%%%%%%%%%%%%%%%%%%%%%%%%%%%%%%%%%%%%%%%%
%%%%%%%%%%%%%%%%%%%%%%%%%%%%%%%%%%%%%%%%%%%%%%%%%%%%%%

The deep inelastic hadronic tensor (for unpolarized scattering) can be defined as 
\begin{equation}
W^{\mu\nu} \, = i \, \int d^4y\, e^{iq\cdot y} \langle P, {\cal Q} \vert \, \Big[ J^\mu (y) , J^\nu (0) \Big] 
\, \vert P, {\cal Q} \rangle \,,
 \label{HadronicTensor}
\end{equation}

\noindent where $ J^\mu(y)$ is the electromagnetic hadron current and $ {\cal Q} $ is the electric charge of the initial  hadron. This tensor can be decomposed into the  structure functions $F_1 (x,q^2) $ and $F_2 (x,q^2) $ as \cite{Manohar:1992tz}
\begin{equation}
W^{\mu\nu} \, = \, F_1 (x,q^2)  \Big( \eta^{\mu\nu} \,-\, \frac{q^\mu q^\nu}{q^2} \, \Big) 
\,+\,\frac{2x}{q^2} F_2 (x,q^2)  \Big( P^\mu \,+ \, \frac{q^\mu}{2x} \, \Big) 
\Big( P^\nu \,+ \, \frac{q^\nu}{2x} \, \Big)
\, ,  \label{Structure}
\end{equation}

\noindent where we use the Minkowski metric $\eta_{\mu\nu}={\rm diag}(-,+,+,+)$. 

As is well known, the cross section for the deep inelastic scattering is related to
the amplitude of forward Compton scattering. This amplitude is determined  by  the tensor 
\begin{equation}
T^{\mu\nu} \, = i \, \int d^4y e^{iq\cdot y} \langle P, {\cal Q} \vert \, {\cal T} \Big(  J^\mu (y) J^\nu (0) \Big)  
\, \vert P, {\cal Q} \rangle\,,
 \label{forwardamplitude}
\end{equation}

\noindent which can be decomposed as 
\begin{equation}
T^{\mu\nu} \, = \, {\tilde F}_1 (x,q^2)  
\Big( \eta^{\mu\nu} 
\,-\, \frac{q^\mu q^\nu}{q^2} \, \Big) 
\,+ \,\frac{2x}{q^2} {\tilde F}_2 (x,q^2)  
\Big( P^\mu \,+ \, \frac{q^\mu}{2x} \, \Big) 
\Big( P^\nu \,+ \, \frac{q^\nu}{2x} \, \Big)
\, ,  \label{CompStructure}
\end{equation}

\noindent where ${\tilde F}_1 (x,q^2)  $ and ${\tilde F}_2 (x,q^2) $ are the associated structure functions. 

The optical theorem relates the tensors $W^{\mu\nu}$ and $T^{\mu\nu}$ and implies that~\cite{Manohar:1992tz}
\begin{equation}
\label{optical}
F_{1,2} (x,q^2) \equiv 2 \pi \,{\rm Im }\,{\tilde F}_{1,2} (x,q^2)\,.
\end{equation}

The imaginary part of the forward Compton scattering amplitude can be expressed in terms of a sum over the intermediate states $X$ with mass $M_X\,$,  formed in the hadron-photon collision
\begin{equation}
\label{Imag}
{\rm Im} T^{\mu\nu} \, = \,  2 \pi^2 \, \sum_X \,\delta \Big( M_X^2 + (P+q)^2 \, \Big) 
\langle P, {\cal Q} \vert J^\nu ( 0 )   \vert P + q,\, X \rangle\,
\langle P + q , \, X \vert J^\mu ( 0 )   \vert P, {\cal Q} \rangle\,\,.
\end{equation}

\subsection{ DIS in the hard wall model}

Polchinski and Strassler found prescriptions for calculating $ {\rm Im} T^{\mu\nu} $ 
from gauge string duality for different regimes of Bjorken parameter $x$~\cite{Polchinski:2002jw} using the hard wall model. 
This model consists of a space $AdS_5 \times W \,$, 
with metric $g_{MN}$: 
\begin{equation}
\label{AdS} ds^2 \equiv g_{MN} \,dx^M dx^N \,= \, \frac{R^2}{z^2}( dz^2  +
\eta_{\mu\nu} dy^\mu dy^\nu  )\,+  \,R^2 ds_W^2\,\,\,, 
\end{equation}

\noindent restricted to the region $ 0 \le z \le 1/\Lambda \,$ , where $\Lambda\,$ is an infrared cut off   interpreted as the QCD scale. This cut off breaks conformal invariance. $W $ is a five dimensional  compact space. 

In the supergravity regime ($x$ of order one) the prescription relates the matrix elements of a hadron $U(1)$  current to a ten dimensional interaction action. 
For the case of the scattering of a scalar particle by a virtual photon with polarization
$ \eta_\mu $ the prescription takes the form  
\begin{eqnarray}
\eta_\mu \langle P_X ,X  \vert {\tilde J}^\mu ( q )   \vert P, {\cal Q}  \rangle\,
&=&  (2 \pi)^4 \, \delta^4 ( P_X - P - q ) \,\eta_\mu \,  \langle P + q, X \vert J^\mu ( 0 )  \vert P,{\cal Q} \rangle\,\nonumber\\
&=& i {\cal Q} \, \int d^{10}x \sqrt{-g} A^m \Big( \Phi_i\partial_m \Phi_X^\ast \,-\, 
  \Phi_X^\ast  \partial_m \Phi_i     \Big)\,\,.
\label{INTERACTION}
\end{eqnarray}

\noindent Here $A_m (x)=(A_z,A_\mu)$ is a Kaluza-Klein gauge field, $ \Phi_i $ and $ \Phi_X$ are the dilaton fields representing the initial and final scalar states. 
The solutions for the free field equations of motion (with the hard cut off condition) were studied in \cite{Polchinski:2002jw} and are represented in terms of the Bessel functions  $K_0 (qz)\,,K_1 (qz)\,$ for the gauge field and $ J_{\Delta-2} ( p z )\,$ for the scalar state with momentum $p$ (that can be $P$ or $P_X $).
Using these solutions in eqs. (\ref{INTERACTION}) and (\ref{Imag})  
they found the structure functions for the scalar case 
\begin{equation}
F_1 (x,q^2)\,=\,0 \,\,\,;\,\,\, F_2 (x,q^2)\,=\, \pi C_0 \, {\cal Q}^2 \left( \frac{\Lambda^2}{q^2} 
\right)^{\Delta - 1} \, x^{\Delta + 1} \, (1- x)^{\Delta - 2}\,,
\end{equation}

\noindent where $C_0$ is a normalization constant and  $\Delta$ is the scaling dimension of the scalar state.

\subsection{ DIS in the soft wall model}

An AdS/QCD phenomenological model that leads to linear Regge trajectories was proposed in \cite{Karch:2006pv}. In this, so called, soft wall model there is an $AdS_5$ space with a static dilaton background field $\varphi $.
In this model there is no hard cut off: $ 0 \le z < \infty \,$.
The infrared cut off is represented by the background dilaton field which is chosen 
as $\varphi = c z^2$. The constant $c$, with dimension of mass squared, is related to the QCD scale.

Inspired in this five dimensional model, here we propose a phenomenological ten dimensional model
represented by bulk actions of the form
\begin{equation}
I \,=\, \int d^{10}x \, \sqrt{-g} \,\, e^{-\varphi} \, {\cal L}\,\,, 
\label{action}
\end{equation}

\noindent where ${\cal L} $ is the lagrangian density  and $g_{MN}$ is  
the ten dimensional metric  of $AdS_5 \times W \,$ space, given in eq. (\ref{AdS})
but now the coordinate $z$ has no hard cut off: $ 0 \le z < \infty \,$. The  ten dimensional dilaton field 
$\varphi $ is also chosen as $\varphi = c z^2$. We will call this model also as soft wall.

So, instead of eq. (\ref{INTERACTION}),  we take the following prescription for the supergravity regime in the presence of the dilaton background 

\begin{eqnarray}
\eta_\mu \langle P_X ,X  \vert {\tilde J}^\mu ( q )   \vert P, {\cal Q}  \rangle\,
&=&  (2 \pi)^4 \, \delta^4 ( P_X - P - q ) \,\eta_\mu \,  \langle P + q, X \vert J^\mu ( 0 )  \vert P,{\cal Q} \rangle\,\nonumber\\
&=& i {\cal Q} \, \int d^{10}x \sqrt{-g}\, e^{-\varphi} \,  A^m \Big( \Phi_i\partial_m \Phi_X^\ast \,-\, 
  \Phi_X^\ast  \partial_m \Phi_i     \Big)\,.\,\,
\label{INTERACTION2}
\end{eqnarray}

It is important to remark that the fields $ A^m $ and $\Phi$ appearing in this equation are not 
the same as those of the hard wall model used in eq. (\ref{INTERACTION}).
This happens because the presence of the dilaton background changes the free field equations of motion. 
In contrast to the hard wall model, where the solutions are represented 
in terms of Bessel functions, in the soft wall we will see that the solutions involve confluent hypergeometric functions  $\,{\cal U} (a;b; cz^2) $ and $\,{\cal M} (a;b;cz^2)$. In fact the normalization condition imposed on the scalar field will reduce the function $\,{\cal M} $ to an associated  Laguerre polynomial $ L_{n}^{m} ( cz^2)\,$ which leads 
to the linear Regge trajectories for the mass spectrum. 
This is different from the asymptotic quadratic trajectories 
obtained in the hard wall model where the masses come from the zeroes of Bessel functions. 

%%%%%%%%%%%%%%%%%%%%%%%%%%%%%%%

{\boldmath
\section{Structure functions in the soft wall for large $x$}}

In order to calculate the deep inelastic scattering structure functions in the soft wall model we have to solve first the equations of motion for the gauge field. The gauge field is a  Kaluza Klein excitation with five components $A_m~=~(A_z, A_\mu)$ that do not depend on the coordinates of the $W$ space. The gauge field 
action in the presence of the soft wall dilaton is 

\begin{equation}
I \,=\,-\,  \int d^{10}x \, \sqrt{-g} \,\, e^{- c z^2 } \, \frac{1}{4} F_{mn} F^{mn} \,\,, 
\end{equation}

\noindent which leads to the equations of motion 
\begin{eqnarray}
\Box  A^\mu  &+& z e^{cz^2}  \partial_z \Big( e^{-cz^2} \frac{1}{z}
\partial_z A^\mu \Big) \,-\,  \eta^{\mu\nu} \partial_\nu \Big( z e^{cz^2} 
\partial_z \Big( e^{-cz^2} \frac{1}{z}
 A_z \Big) \,+\, \partial_\rho A^\rho \Big) \,=\,0 \nonumber\\
\Box A_z &-&  \partial_z \Big( \partial_\mu A^\mu \Big) \,=\, 0\,.
\end{eqnarray} 

\noindent We use the notation:  $A^\mu \equiv \eta^{\mu\nu} A_\nu\,$ and $\Box \equiv \eta^{\mu\nu}\partial_\mu \partial_\nu\,$. 

In order to solve these equations of motion we choose the gauge condition
\begin{equation}
 \partial_\rho A^\rho \,+\,
z e^{cz^2} \partial_z \Big( e^{-cz^2} \frac{1}{z}
 A_z \Big) \,=\,0 \,,
\end{equation}

\noindent and impose that the boundary value of the gauge field represents a virtual photon with polarization $\eta^\mu$ and  momentum $ q^\mu$ 
\begin{equation}
A_\mu (z, y) \vert_{z\to 0} \,=\, \eta_{\mu} \, e^{iq\cdot y} \,,
\end{equation}

\noindent where $q\cdot y \equiv q^\mu y_{\mu} \,$ and   $q^2 = q^\mu q_{\mu} \,>\, 0\,$.
The corresponding solutions are
\begin{eqnarray}
A_\mu (z, y) &=& \eta_\mu \, e^{iq\cdot y} \,c \,\,  \Gamma (1+\frac{q^2}{4c} )\,\, z^2 \,\,{\cal U} (1+\frac{q^2}{4c} ; 2 ; cz^2 )
\nonumber\\
A_z (z, y)  &=& \frac{i}{2} \,  \eta \cdot q \,  e^{iq\cdot y} \,\, \Gamma (1+\frac{q^2}{4c} )\,\, z \,\, 
{\cal U} (1+\frac{q^2}{4c} ; 1 ; cz^2 )\,, 
\label{Gauge}
\end{eqnarray} 

\noindent where $\,{\cal U} (a;b;w) \,$ are the confluent hypergeometric functions of the second kind. We note that both products $ \, {\cal U} (a;2;w) \,w \,  \Gamma (a) \,$ and $\, {\cal U} (a;1;w) \, \sqrt{w}\, \Gamma (a) \,$ decrease rapidly for $ a \, w > 1$. So it is natural to define an effective maximum value for the radial coordinate 
\begin{equation}
z_{int} \approx \frac{1}{\sqrt{c\,(1+\frac{q^2}{4c})\,}}\,\sim \, \frac{1}{q}\,\,,
\end{equation}

\noindent independent of the infrared cut off scale $c$. For $z > z_{int}$ the gauge field becomes very small so that the interaction between the photon and the hadron is negligible.  
Based on this fact we will make the approximation that the interaction occurs only at $z \le z_{int}\,$. 
Note that this is not an infrared cut off in the space. There is no boundary condition at $z = z_{int}\,$.

The four dimensional center of mass energy squared ${s}=-{P_X}^2\approx q^2( \frac{1}{x} \,-\,1\,)$ is holographically related to the ten dimensional energy scale ${\tilde{s}}$ by

\begin{equation}
\tilde s \,\le \, \frac{z^2}{R^2} \, s \,. 
\end{equation}

\noindent Using the approximation that the interaction occurs at  $ z \le   z_{int} \,$ we have 
\begin{equation}
\tilde s \,\le \,  
\frac{z_{int}^2}{R^2} \,q^2 \, 
\Big( \frac{1}{x} \,-\,1\,\Big) \,< \, 
\frac{1}{\alpha'\,( 4 \pi g N)^{1/2}} \,\frac{1}{x} \,.
\end{equation}

The supergravity approximation used in this section is valid when the ten dimensional energy is not sufficient to produce massive states. This corresponds to $\alpha' \, {\tilde s}\, < 1$, so we must have $ x > > (gN)^{-1/2}\,$. 
Then, the large $x$ regime considered in this section corresponds to: 
$ 1 >  x > > (gN)^{-1/2}\,$.

%%%%%%%%%%%%%%%%%%%%%%%%%%%%%%%%%%%%%%%%%%%%%%%%%%%%%%%%%%%%%%%%%

\vskip .5cm

In order to calculate the structure functions for scalar particles we need to solve the corresponding equations of motion. In this case, the soft wall action (\ref{action}) involves a ten dimensional Lagrangian density 
${\cal L} \, = \partial_M \Phi \partial^M \Phi  \,$.  Since the space is a direct product of $AdS_5 $ and $W$ it is convenient to decompose the ten 
dimensional scalar field as
\begin{equation}
\Phi (z, y, \Omega ) \, = \, \phi (z, y) \, Y ( \Omega )\,\,,
\end{equation}

\noindent where $ \Omega$ are the angular coordinates of the space $ W$. Assuming that $ Y (\Omega ) $ is an eigenstate of the Laplacian in the coordinates $\Omega$, the ten dimensional equation reduces to 
 \begin{equation}
z^3  e^{cz^2} \partial_z \Big( e^{-cz^2} \frac{1}{z^3} \partial_z \Phi \Big) \,+\, \Box  \Phi \,-\, 
\frac{R^2}{z^2} \,{m_5}^2\, \Phi \,=\,0 \,,
\end{equation}

\noindent where $m_5$ is related to the eigenvalues of $Y (\Omega )$.
The scaling dimension $\Delta$ of the boundary operator  is related to $m_5$ in the AdS/CFT correspondence by $ \Delta \,=\, 2 + \sqrt{ 4 \,+\, {m_5}^2 R^2}$.

The solution that is normalizable, taking a plane wave for  the $y^\mu $ coordinates,
representing a particle with momentum $p$, is

\begin{equation}
 \Phi \,=\, d \, e^{ip \cdot y}\, z^\Delta\, {\cal M} ( \frac{p^2}{4c} 
+ \frac{\Delta}{2}; \Delta - 1; cz^2 \,)\,Y ( \Omega )\,\,,
\end{equation}

\noindent where ${\cal M} ( a ; b ; u) \,$ is the confluent hypergeometric function of the first kind and $d$ is a normalization constant.
We impose the normalization condition for the radial and angular coordinates $ (z , \Omega^i\,) \, $ in the soft wall background
\begin{equation}
\label{norm}
\int dz \,d^5 \Omega \,\,\frac{R^8}{z^3}\,\sqrt{g_W}\, e^{-cz^2} \,\vert \, \Phi\, \vert^2 \,=\,
\, R^8\,\, \int \frac{dz}{z^3} \, 
e^{-cz^2} \,\vert \,\phi (z,y)\,  \vert^2  \,=\, 1\,\,,
\end{equation}

\noindent where we have used the angular normalization
\begin{equation}
\int d^5 \Omega \,\,\sqrt{g_W}\,\vert Y (\Omega ) \,\vert^2 \, = \,1  \,\,.
\end{equation}

 The normalization condition (\ref{norm}) can only be satisfied if the first argument of the confluent hypergeometric function is a non-positive integer:
\begin{equation}
\frac{p^2}{4c} + \frac{\Delta}{2}\,=\, -n\,\,.
\end{equation}

\noindent  Identifying $p^2 \,=\, -{m_n}^2\,$ we see how the soft wall leads to a discrete mass spectrum with linear Regge trajectories for normalizable modes
\begin{equation}
\label{discretemass}
{m_n}^2 \,=\, 4 c  \Big( n +  \frac{\Delta}{2}\,\Big)\,,
\end{equation}

\noindent where $ \sqrt{c} $ represents an infrared cut off mass scale. 
The confluent hypergeometric function reduces then to an associated Laguerre polynomial $L^m_n (u)$, so that  the normalized solution reads
\begin{equation}
 \Phi_n (y, z, \Omega)\,  = \, \Big[ \frac{2 c^{\Delta-1} \Gamma( n + 1) }{  \Gamma (n + \Delta -1 ) }\Big]^{1/2}\,
\frac{1}{R^4} \, e^{ip \cdot y}\, z^\Delta L^{\Delta - 2}_n\, ( cz^2\,)\,Y ( \Omega )\,\, .
\label{scalar}
\end{equation}

For the initial scalar state we choose a field $\Phi_i $ with momentum  $p = P$  
and $ n = 0$ corresponding  the lowest mass in the spectrum
\begin{equation}
\Phi_i \,\equiv \Phi_i(y,z,\Omega) \, =\, \Big[ \frac{2 c^{\Delta-1} }{  \Gamma ( \Delta -1 ) }\Big]^{1/2}\,
\frac{1}{R^4} \,\, e^{i P \cdot y}\, z^\Delta \, \,Y ( \Omega )\,\,.
\label{initialstate}
\end{equation}

For the final scalar state we take a field $ \Phi_X$ with momentum $p = P_X $ 
so that  
\begin{equation}
\label{n}
n \,=\, n_X \,=\, - \frac{ P_X^2}{4c} \,-\, \frac{\Delta }{2} \,=\, \frac{s}{4c} \,-\, \frac{\Delta }{2}\,,
\end{equation}

\noindent where we have used momentum conservation $P_X = P  + q\,$. This implies that this state is given by
\begin{equation}
\Phi_X \,\equiv \,\Phi_X (y , z, \Omega) \, =\, \Big[ \frac{2 c^{\Delta-1} 
\Gamma( \frac{s}{4c} \,-\, \frac{\Delta }{2} + 1) }
{  \Gamma (\frac{s}{4c} \,+\, \frac{\Delta }{2} -1 ) }\Big]^{1/2}\,
\frac{1}{R^4} \,\, e^{i P_X  \cdot y}\, z^\Delta L^{\Delta - 2}_{ n_X }
\, ( cz^2\,)\, Y ( \Omega )\,\,
\label{finalstate}
\end{equation}

Now we perform the soft wall version of the integral  (\ref{INTERACTION}) representing the interaction amplitude. We use the  solutions (\ref{Gauge}) for the gauge field, and
(\ref{initialstate}), (\ref{finalstate}) for the scalars. The result is
\begin{eqnarray}
&i& {\cal Q} \int d^{10}x \sqrt{-g} e^{-\varphi} \, A^m \Big( \Phi_i\partial_m \Phi_X^\ast \,-\, 
  \Phi_X^\ast  \partial_m \Phi_i     \Big)\nonumber\\
&=& {\cal Q} (2 \pi )^4 \delta^4 (P+ q - P_X ) \, 2  \,\eta_\mu \Big[ P^\mu + \frac{q^\mu}{2x} \Big] 
\Big( \frac{ \Gamma (\frac{s}{4c} - \frac{\Delta}{2} +1 )}
{\Gamma (\Delta - 1 ) \Gamma (\frac{s}{4c} + \frac{\Delta}{2} - 1 )} \Big)^{1/2} \times \nonumber\\
& & \Gamma ( 1 + \frac{q^2}{4c} )\,
\int_0^\infty  dw \, w^{\Delta - 1}\,\, e^{-w}  
\,\,{\cal U}  ( 1 + \frac{q^2}{4c} ; 2 ; w ) \,\,
L^{\Delta - 2}_{n_X} (w) \nonumber\\ 
&=& (2 \pi )^4 \delta^4 (P+ q - P_X ) \, 2\,{\cal Q} \,\eta_\mu \Big[ P^\mu + \frac{q^\mu}{2x} \Big] 
\,\frac{q^2}{4c} \, \,( \Delta - 1) \, \Big[  \Gamma (\Delta - 1) \, \Big]^{1/2}\,
f(q,s)\nonumber\\ 
\label{scalarint}
\end{eqnarray}

\noindent where $ w = cz^2\,$ and we defined 
\begin{eqnarray}
\label{f(q,s)}
f(q,s) \equiv \Big[ \frac{\Gamma ( \frac{s}{4c} + \frac{\Delta}{2} -1 ) } {\Gamma( \frac{s}{4c} -  \frac{\Delta}{2} + 1 ) } \Big]^{1/2}\,\,
\frac{ \Gamma ( \frac{q^2}{4c} +\frac{s}{4c}  - \frac{\Delta}{2}  ) }
{\Gamma( \frac{q^2}{4c} + \frac{s}{4c} +  \frac{\Delta}{2} ) }\,.
\end{eqnarray}
 
\noindent The integral over $w$ in eq. (\ref{scalarint}) was calculated using an integral representation for the $ {\cal U }\,$ function. Substituting this result in (\ref{INTERACTION}) we find the matrix elements of the current 
\begin{eqnarray}
\label{Smatrixelem}
 \langle P + q, X \vert J^\mu ( 0 )  \vert P,{\cal Q} \rangle 
\,=\,  2\,{\cal Q} \, \Big[ P^\mu + \frac{q^\mu}{2x} \Big] 
\frac{q^2}{4c} \,( \Delta - 1 ) \, \Big[ \Gamma (\Delta -1 ) \Big]^{1/2}\, f(q,s)\,\,.\nonumber\\ 
\end{eqnarray}

Then, the imaginary part of the forward Compton scattering amplitude in (\ref{Imag}) reads 
\begin{eqnarray}
{\rm Im} T^{\mu\nu} &=&  8 \pi^2 \,{\cal Q}^2 \, \sum_X \,\delta \Big( M_X^2 + (P+q)^2 \, \Big) 
\Big[ P^\mu + \frac{q^\mu}{2x} \Big]\, \Big[ P^\nu + \frac{q^\nu}{2x} \Big] 
 \,\Big( \frac{q^2}{4c}\,\Big)^2\,\nonumber\\
&\times& (\Delta - 1)\,\Gamma ( \Delta )\,  \Big[ f(q,s)\Big]^2\,\,.
\end{eqnarray}

From equation (\ref{discretemass}) for the soft wall we see that the spacing between the masses $m_X$ is small compared with $q$ so that the sum over the states $X$ can be approximated by an integral 
\begin{equation}
\label{nova1}
\sum_X \,\delta \Big( M_X^2 + (P+q)^2 \, \Big) 
\,=\,  \frac{1}{4c}\,\, \int dn  \,\delta \Big( n - \frac{s}{4c}  + \frac{\Delta}{2} \Big)\,
=\,\frac{1}{4c}\, .
\end{equation}

\noindent So, from  eqs. (\ref{CompStructure}) and (\ref{optical}) we find 
\begin{eqnarray}
F_1 \,=\, 0 \;\;;\qquad 
F_2 \,=\, 8 \pi^3 \,\frac{{\cal Q}^2}{x} \,(\Delta - 1) \,\Gamma (\Delta ) \,
\Big( \frac{q^2}{4c}\Big)^3 \, \Big[ f(q,s)\Big]^2 \,\,.
\label{F2escalar}
\end{eqnarray}

\noindent These are the structure functions for scalar states in the soft wall model.
This is our main result for the scalar case in the supergravity regime. In order to compare this result
with the one obtained in the hard wall model we are going to consider evaluate this structure function at leading order in $c/q^2\,$. Note that in the deep inelastic scattering limit: 
$ q^2/4c >> 1\,$ with $x$ fixed. So we have

\begin{equation}
\frac{q^2}{4c} \Big( \frac{1}{x} - 1 \Big) \,>>\,1 \,\, .
\end{equation}

\noindent Using this approximation in the relation
\begin{equation}
\frac{s}{4c}\,=\, -\frac{p^2}{4c} \,+\, \frac{q^2}{4cx}\,-\, \frac{q^2}{4c}\,=\, \frac{\Delta}{2}\,+\,\,
\frac{q^2}{4c} \Big( \frac{1}{x} - 1 \Big) \,\,,
\end{equation}

\noindent we find that the ratios of the gamma functions in eq. (\ref{f(q,s)}), at leading order in 
$c/q^2\,$, reduce to 
\begin{equation}
\frac{ \Gamma ( \frac{s}{4c} + \frac{\Delta}{2} -1 ) }
{\Gamma( \frac{s}{4c} -  \frac{\Delta}{2} + 1 ) } \,\approx \, 
\Big[ \,\frac{q^2}{4c} \Big( \frac{1}{x} - 1 \Big) \,\Big]^{\Delta - 2}
\,\,\,\,\,;\,\,\,\,\,\,
\Big[\,\frac{ \Gamma ( \frac{q^2}{4c} +\frac{s}{4c}  - \frac{\Delta}{2}  ) }
{\Gamma( \frac{q^2}{4c} + \frac{s}{4c} +  \frac{\Delta}{2} ) }\,\Big] 
\,\approx \,\,
\Big[ \,\frac{q^2}{4cx}\,\Big]^{\,-\Delta }\,.
\end{equation}

\noindent So the structure function $F_2$  reads
\begin{equation}
F_2 \,\approx \, 8 \,\pi^3\, {\cal Q}^2 (\Delta - 1) \,\Gamma (\Delta ) \, 
\Big( \frac{4c}{q^2}\Big)^{\Delta -1} \,(1 - x)^{\Delta - 2} \, x^{\Delta + 1}\,\,.
\end{equation}

\noindent This leading order result coincides (up to numerical factors) with the scalar structure functions found in ref.\cite{Polchinski:2002jw} using a hard cut off condition, when  we identify the soft and hard wall mass scales $\sqrt{c}$ and $\Lambda$.

It is interesting to observe that the matrix element of the hadronic current obtained in eq. 
(\ref{Smatrixelem}) holds for any value of $q^2 $ and $x$ (as long as  $ (gN)^{-1/2} << x < 1 \,$).
In particular in the elastic limit $ x \to 1 $ this matrix element is related to the elastic 
form factor of a scalar particle by 

\begin{equation}
\label{} \lim_{x\to 1} \langle P + q, X \vert J^\mu ( 0 )  \vert P,{\cal Q} \rangle 
\,=\, 2 ( 2 P + q )^\mu \, F (q^2 ) \,.
\end{equation}

So we find the soft wall scalar form factor 

\begin{equation}
\label{Formfactor}
F (q^2 ) \,=\, \frac{{\cal Q}}{2} \, \Gamma (\Delta ) 
\frac{ \Gamma ( \frac{q^2}{4c} + 1  ) }
{\Gamma( \frac{q^2}{4c} + \Delta ) }\,.
\end{equation}

This result was obtained previously in ref. \cite{Brodsky:2007hb}. 
In this reference,  it was shown that for the pion ($\Delta = 2 $) the 
soft wall form factor is in better agreement than the hard wall form factor 
when compared with results obtained from experimental data.

%%%%%%%%%%%%%%%%%%%%%%%%%%%%%%%%%%%%%%%%%%%%%%%%%%%%%%%%%%%%%%%%%

%%%%%%%%%%%%%%%%%%%%%%%%%%%%%%%%%%%%%%%%%%%%%%%%%%%%%%%%%%%%%%%%%%%%%%%%%%%%%%%%

{\boldmath
\section{ Structure functions at small $x$ }}

In the previous section we calculated the deep inelastic scattering amplitudes in the case $ (gN)^{-1/2} << x < 1 \,$. In that case we used supergravity approximation for string theory since the ten dimensional energy scale $\sqrt{\tilde s}$ was not high enough to excite massive string modes. Now we will consider  a regime of small $x$ corresponding to $ \exp \,( - \sqrt{gN }) << x << (gN)^{-1/2}\,$. In this case there are  massive string excitations so that we should, in principle, consider string scattering amplitudes in $AdS_5 \times W$ space. However, the condition $ \exp \,( - \sqrt{gN }) << x $ implies that the strings are small compared to the AdS radius and we can  approximate locally the amplitudes by those of flat space.  

Now the four dimensional forward scattering amplitude 
\begin{equation}
 \eta_\mu \eta_\nu \, T^{\mu\nu} \,(2\pi)^4 \,\delta^4 ( q - q' ) \,
\end{equation}

\noindent will be identified with the ten dimensional string amplitude \cite{Polchinski:2002jw}. For the soft wall the string amplitude is 
\begin{eqnarray}
S_{10} &=& \int d^{10}x \sqrt{-g} \,
e^{-cz^2}  \,\,\Big( {\cal K} \, G \, \Big)\vert_{_{t=0}}\,   \nonumber\\
&=& \frac{1}{8} \int d^{10}x \sqrt{-g} \,e^{-cz^2}\,
\Big\{ 4 v^a v_a \partial_m \Phi  F^{mn} F_{pn} \partial ^p \Phi \nonumber\\
&-& \Big( \partial^{M} \Phi \partial_{M} \Phi  v^a v_ a \,+\, 2 v^a \partial_a \Phi v^b \partial_b \Phi\, \Big)
F_{mn} \,F^{mn} \Big\} \, G\vert_{_{t=0}} \,\,,
\label{S_10}
\end{eqnarray}

\noindent where $v^a$ are the Killing vectors of the compact $W$ space.
In this expression ${\cal K}$ represents a ten dimensional kinematic factor, where each field  represents one of the four interacting string states associated with the four dimensional particles.
The factor $G$ is a flat space Veneziano amplitude
\begin{equation}
G\,=\, \frac{{\alpha^\prime}^3{\tilde s}^2}{64}\, \prod_{{\tilde \xi}= {\tilde s},{\tilde t},{\tilde u}} 
\frac{\Gamma(-{\alpha^\prime}{\tilde \xi}/4)}{\Gamma(1+{\alpha^\prime}{\tilde \xi}/4)}
\end{equation}

\noindent to be evaluated at $t\equiv p^\prime-p=0$ which represents a four dimensional forward scattering. The ten dimensional Mandelstam variables $\,\tilde t \,,\,\tilde s\,$ are related to the four dimensional variables
$ t , s$ by

\begin{equation} 
\alpha' \tilde s \,=\, \alpha' s \, \frac{z^2}{R^2} \,+\, \frac{\alpha' }{R^2} \,
\Big( - 3\, z \partial_z \,+\, z^2 \partial^2_z \,+\, \nabla^2_W \Big) \,
\end{equation} 
 
\begin{equation} 
\alpha' \tilde t \,=\, \alpha' t \, \frac{z^2}{R^2} \,+\, \frac{\alpha' }{R^2} \,
\Big( - 3\, z \partial_z \,+\, z^2 \partial^2_z \,+\, \nabla^2_W \Big) \,
\end{equation} 
 
\noindent So that, for the forward scattering condition $\, t =0\,$, 
$\, \alpha' \tilde t \,$ does not vanish because it contains contributions from the radial and angular momenta which are of order of $\,(gN)^{-1/2}\,$. Since ${\cal K}$ is real, the imaginary part of $S_{10}$ is related to the imaginary part of $G$ which at $ \, t = 0 \,$ is
\begin{equation} 
{\rm Im}\,G\vert_{_{t=0}}\,=\, \frac{\pi \alpha' }{4}
\sum_{\ell=1}^{\infty} \, \delta ( \ell - \frac{\alpha' \tilde s }{4} \,) 
\,(\ell)^{\,\alpha' \tilde t /2}\,\,.
\end{equation}

\noindent The factor $ (\ell)^{\,\alpha' \tilde t /2}$ can be approximated using the delta function and

\begin{equation}
(\alpha' \tilde s\,)\,\approx\, \alpha' \,\frac{z^2}{R^2} \,\,s \,\,<< \frac{1}{x}\,\,. 
\end{equation}

We have that $(\alpha' \,\tilde s\,)^{\alpha' \tilde t /2} \sim 1$ when $ \exp (\, - \sqrt{gN}\, )  << x $. Thus

\begin{equation} 
{\rm Im}\,G\vert_{_{t=0}}\,\approx \, \frac{\pi \alpha' }{4}
\sum_{\ell=1}^{\infty} \, \delta ( \ell - \frac{\alpha' \,s \, z^2  }{ 4 \,R^2}  \,) \,\,.
\end{equation}

In the kinematic factor $\cal K$, the field strength  $F_{mn}$ is associated with an incoming 
photon of four momentum $q_\mu$ and an outgoing photon of momentum $ q^\prime_\mu $
while $\Phi$  represents the incoming and outgoing scalar states with four momentum $P_\mu$. 
These fields are represented by the supergravity solutions given in the previous sections involving four dimensional plane waves. The derivatives $\partial_\mu$ acting on these solutions generate the corresponding four dimensional momenta.   The condition $x << 1$ implies that $P \cdot q >> q^2 >> P^2 $, so that the dominant term in $\cal K$ will be the one corresponding to  $(\,P \cdot q\, )^2$.  This contribution comes from the first term in eq. (\ref{S_10}), with 
$m = \mu\,$ and $ p = \nu$. Then we have
\begin{eqnarray} 
{\rm Im} \, S_{10} &=& \frac{\pi \alpha' }{8}
\sum_{\ell=1}^{\infty} \, \int d^{10}x \sqrt{-g} \,
e^{-cz^2}  \,v^a v_a \nonumber\\
&&\times \,\partial_{\mu} \Phi (-P) \,\partial^{\nu }\Phi (P)\,
  F^{\mu n} (-q' ) \,  F_{\nu n} (q)   \,  
\, \delta ( \ell \,-\, \frac{\alpha' s z^2}{ 4R^2} \,)\,.
\label{Acaointeracao}
\end{eqnarray}

The field strengths,  calculated from the solutions in eq. (\ref{Gauge})
for the gauge field are
\begin{eqnarray}  
F_{0\mu} (q) &=& \frac{z}{2}\,  e^{i q \cdot y}\,
 \Big[ \,q_{\mu} (q \cdot \eta ) \,-\, \eta_{\mu} q^2 \,\Big]\,
\Gamma (1+ \frac{q^2}{4c}) \,\,
{\cal U} ( 1+ \frac{q^2}{4c} ; 1; cz^2) 
\nonumber\\
F_{\mu\nu} (q) &=& i\, cz^2\, e^{i q \cdot y} \,
\Big[ \,q_{\mu} \eta_\nu \,-\, q_\nu \eta_\mu \Big] \,\Gamma (1+ \frac{q^2}{4c})\,\,
{\cal U} ( 1+ \frac{q^2}{4c} ; 2; cz^2) \,.
\label{termovetorial}
\end{eqnarray} 
 
For the scalar states, using the solution for the initial state in eq. (\ref{initialstate}) we find
\begin{equation}
\partial_\mu \Phi (-P) \partial^\nu \Phi (P) \,=\, 
P_\mu P^\nu
 \frac{2 c^{\Delta-1} }{  \Gamma ( \Delta -1 ) }\,
\frac{z^{2\Delta + 2}}{R^{10}} \,  \, 
\,\vert \,Y ( \Omega )\,\vert^2 \,.
\label{termoescalar}
\end{equation}

The angular normalization integral is 
\begin{equation}
\int d^5 \Omega \,\sqrt{g_W} v^a v_a 
\,\vert \,Y ( \Omega )\,\vert^2 \,
=\, \rho \, R^2\,,
\end{equation}

\noindent where $\rho$ is some dimensionless quantity.

Using the results (\ref{termovetorial}) and (\ref{termoescalar}) in the interaction action (\ref{Acaointeracao}) and integrating over $ y $ and $ \Omega $ we find
\begin{eqnarray} 
{\rm Im} S_{10} &=& (2\pi)^4 \delta^4 (q - q' )\,\frac{\pi \alpha' \,\rho}{8\,R^{2}}\, 
\frac{2 c^{\Delta-1} }{  \Gamma ( \Delta -1 ) }\,
P_\mu P^\nu 
\sum_{\ell=1}^{\infty} \, \int dz \,e^{-cz^2} 
\,z^{2\Delta + 3\,}\,
\,\,\Gamma^2 (1+ \frac{q^2}{4c})\,\,
\nonumber\\
&\times & \Big\{ \frac{1}{4}\,\big[ \,q^{\mu} (q \cdot \eta ) \,-\, \eta^{\mu} q^2 \,\big]\,
\big[ \,q_{\nu} (q \cdot \eta ) \,-\, \eta_{\nu} q^2 \,\big]\,
\,{\cal U}^2 ( 1+ \frac{q^2}{4c} ; 1; cz^2) \nonumber\\
&+& c^2\,z^2\, \big[ \,q^{\mu} \eta^\gamma \,-\, q^\gamma \eta^\mu \big]
\big[ \,q_{\nu} \eta_\gamma \,-\, q_\gamma \eta_\nu \big]
 \,\,{\cal U}^2 ( 1+ \frac{q^2}{4c} ; 2; cz^2) \,\Big\}  
\, \delta ( \ell \,-\, \frac{\alpha' s z^2}{ 4R^2} \,)\,.\nonumber\\
\label{Acaointeracao2}
\end{eqnarray}

We can write  the delta function as
\begin{equation}
\delta ( \ell \,-\, \frac{\alpha' s z^2}{ 4R^2} \,)\,
=\,\frac{2R^2}{\alpha'  s z_\ell }\,\delta (z - z_\ell )\,,
\end{equation}

\noindent where 
\begin{equation}
 z_\ell\,=\, 2R \, \sqrt{\frac{\ell}{\alpha' \,s\,}}\, \approx \, \frac2q \, ( 4 \pi \, g N ) ^{1/4} \, (\ell \, x)^{1/2}\, .  
\end{equation}

After integrating over $z$ and identifying the 10-d string amplitude with the 4-d amplitude we find that  
\begin{equation}
{\rm Im} T^{\mu\nu} \,=\, 
\frac{\pi\,\rho\,c^{\Delta-1}}{ 8  \Gamma (\Delta - 1 )\,}\,\,
\frac{(q^2)^2}{s\,x^2}
\Big\{\,\,\big[ \eta^{\mu\nu} - \frac{q^\mu q^\nu}{q^2} \big] \,{\cal A}_2
+\, \big[ \,P^\mu + \frac{ q^\mu}{2x}\,\big] 
\,\,\big[ \,P^\nu + \frac{ q^\nu}{2x}\,\big]\, 4x^2\,
\big( {\cal A}_1\,+\, \frac{{\cal A}_2}{q^2}\,\big)\,\,\Big\} \label{imT}\,,
\end{equation}

\noindent where we have defined 
\begin{eqnarray}
 {\cal A}_1 &\equiv& \frac{1}{4}\,\Gamma^2 ( a )\,
\sum_{\ell=1}^{\infty} \, e^{-cz_\ell^2} \,z_{\ell}^{2\Delta + 2}\,
{\cal U}^2 ( a ; 1; c z_{\ell}^2) 
\nonumber\\
  {\cal A}_2 &\equiv& c^2\, \Gamma^2 (a )\,
\sum_{\ell=1}^{\infty} \, e^{-cz_\ell^2} \,z_{\ell}^{2\Delta + 4}\,
{\cal U}^2 ( a ; 2; c z_{\ell}^2) \,,
\end{eqnarray}

\noindent with $ a\,=\, 1+ \frac{q^2}{4c}\,$. So we obtain from eq (\ref{imT})
\begin{eqnarray}
F_1  &=&  \frac{\pi^2 \,\rho\,c^{\Delta-1}}{ 4  \Gamma (\Delta - 1 )\,}\,\,
\frac{(q^2)^2}{s\,x^2} \, {\cal A}_2 \, \nonumber \\
F_2  &=&  \frac{\pi^2 \,\rho\,c^{\Delta-1}}{ 4  \Gamma (\Delta - 1 )\,}\,\,
\frac{(q^2)^2}{s\,x^2} \, (2 \, x \, q^2 ) \, \big( {\cal A}_1\,+\, \frac{{\cal A}_2}{q^2}\,\big) \,.
\label{f1f2smallx}
\end{eqnarray}

\noindent These are the soft wall structure functions for the small $x$  regime. This is the main result 
of this section. In order to evaluate these structure functions at leading order in $c/q^2$ we define 
 $\,\,\zeta_{\ell}\,\equiv \, (a-1) c z_{\ell}^2 \,\,$ so that  
\begin{equation}
\zeta_{\ell}\,=\, (a-1) c z_{\ell}^2 \,=\, \frac{q^2}{4} \,z_{\ell}^2 \,< \, q^2 \,z_{int}^2 \approx 1\,.
\end{equation}

\noindent Then we can consider $\zeta_{\ell}$ to be bounded in the deep inelastic limit 
and we can use 
\begin{equation}
\lim_{a\to \infty} {\cal U} ( a ; b ; \frac{\zeta}{a-1})\,=\, \frac{2}{\Gamma ( 1 + a - b) } \,
\zeta^{(1-b)/2}
\,K_{b-1} (2\sqrt{\zeta} )  \,,
\end{equation}

\noindent and 
\begin{equation}
\lim_{a\to \infty} e^{-cz_\ell^2}\,=\, \lim_{a\to \infty} e^{-\frac{\zeta_\ell}{a-1}} \,= \, 1\,.
\end{equation}

So the series reduce to
\begin{eqnarray}
 {\cal A}_1 &\approx& \Big[ \frac{q^2}{4} \Big]^{-\Delta - 1} \,
\sum_{\ell=0}^{\infty} \, \zeta_\ell^{\Delta + 1} \,
K_0^2 (2\sqrt{\zeta_\ell} ) 
\nonumber\\
  {\cal A}_2 &\approx&  4 \, \Big[ \frac{q^2}{4} \Big]^{-\Delta } \,
\sum_{\ell=0}^{\infty} \, \zeta_\ell^{\Delta + 1} \,
K_1^2 (2\sqrt{\zeta_\ell} ) \label{series}\,\,,
\end{eqnarray}

\noindent where we have included null $\ell = 0$ terms. 
These series can be approximated by integrals since 
$\zeta_{\ell + 1} - \zeta_\ell \,=\, \sqrt{\,4\pi g N\,}\,x\,<< 1$.
Defining $ \omega \equiv 2 \,
 \sqrt{\zeta} $, we obtain 
\begin{eqnarray}
{\cal A}_1 &\approx&  \, \frac{\left (  q^2 \right )^{-\Delta-1}}{2 \, x \, (4 \pi \, gN)^{1/2}} \, {\cal I}_{\, 0 , \, 2\Delta+3} \nonumber \\
{\cal A}_2 &\approx& \frac{\left (  q^2 \right )^{-\Delta}}{2 \, x \, (4 \pi \, gN)^{1/2}} \, {\cal I}_{ \, 1 , \, 2\Delta+3} \, , 
\label{finala1a2}
\end{eqnarray}

\noindent where 
\begin{equation}
{\cal I }_{\, j , \, n } \equiv   \int_0^{\infty} d \omega  \, \omega^n \, K_{j}^2(\omega) \,
= \, 2^{n-2}\frac{\Gamma(\frac{n+1}{2}+j) \,  \Gamma(\frac{n+1}{2}- j ) \, \Gamma^2(\frac{n+1}{2})}{\Gamma ( n+1) } \, \, . 
\end{equation}

From (\ref{f1f2smallx}) and (\ref{finala1a2}) we have that 
\begin{eqnarray}
F_1  &\approx&  \frac{\pi^2 \,\rho}{ 8 \, (4 \pi \, gN)^{1/2} \, \Gamma (\Delta - 1 )\,}\,\frac{1}{x^2}   \,
\left(\frac{c}{q^2}\right)^{\Delta-1} \,  {\cal I}_{ \, 1 , \, 2\Delta+3}\, \nonumber \\
F_2  &\approx&  2 \, x \, F_1  \, \frac{{\cal I}_{ \, 0 , \, 2\Delta+3}+ {\cal I}_{ \, 1 , \, 2\Delta+3}}{{\cal I}_{ \, 1 , \, 2\Delta+3}} \, = \, 2 \, x \, \frac{2 \Delta + 3}{\Delta +2} \, F_1 \, \, , 
\label{finalf1f2smallx}
\end{eqnarray}

\noindent where we have used the relation ${\cal I}_{ \, 0 , \, n} = \frac{n-1}{n+1} \, {\cal I}_{ \, 1 , \, n}$ \, . As in the large $x$ case of section {\bf 3}, the soft wall deep inelastic structure functions for small $x$  are in agreement at leading order with the hard wall structure functions \cite{Polchinski:2002jw}.

\vskip .5cm 

%%%%%%%%%%%%%%%%%%%%%%%%%%%%%%%%%%%%%%%%%%%%%%%%%%%%%%%%%%%%%%%%%%%%%%%%

\section{Conclusions}

In this article we have calculated the deep inelastic structure functions at large $gN$ using a  phenomenological ten dimensional soft wall model. We investigated two different regimes of the Bjorken parameter $x$: $ (gN)^{-1/2} << x < 1 \,$ and  $ \exp \,( - \sqrt{gN }) << x << (gN)^{-1/2}\,$. In the first regime we performed a supergravity calculation and in the second regime we considered the contribution of massive string states. We found that at leading order the structure functions for soft and hard wall models are the same.  This result could be expected since high energy processes are mapped in  the small $z$ UV region of AdS space while the hard and soft wall models differ mostly in the large $z$ IR region. 
However the calculation of the structure functions involve the mass spectrum and the free field solutions 
which are different in these models. So, there has to be some non trivial compensation that leads to
the same result.
 
In the soft wall model the Regge trajectories for mesons are linear while for the hard wall they are quadratic. As a consequence, the contributions from the sum over intermediate states to 
the structure functions in eq. (\ref{Imag}) are not the same.
The mass spectrum of the final hadronic states implies that in the soft wall model  
\begin{equation}
\label{nova2}
\sum_X \,\delta \Big( M_X^2 + (P+q)^2 \, \Big) 
\,=\,  \frac{1}{4c}\, .
\end{equation}

\noindent  This relation differs from that found in the hard wall model \cite{Polchinski:2002jw}: 

\begin{equation}
\label{nova3}
\sum_X \,\delta \Big( M_X^2 + (P+q)^2 \, \Big) 
\,=\, \frac{1}{2\pi s^{1/2} \,\Lambda\,}\,.
\end{equation}

The fact that the field solutions are different implies also a difference in the hadronic currents. 
It is straightforward to show from eq. (\ref{Smatrixelem}) and ref. \cite{Polchinski:2002jw}
that in the DIS limit ($q^2 \to \infty $ with $x$ fixed), 
the matrix elements of the hadronic currents in the soft wall (SW) and hard wall (HW) models are related by

\begin{equation}
\label{SmatrixComp}
 \langle P + q, X \vert J^\mu ( 0 )  \vert P,{\cal Q} \rangle_{HW} 
\,\sim \, \Lambda^{-1/2} \,s^{1/4}\,    \langle P + q, X \vert J^\mu ( 0 )  \vert P,{\cal Q} \rangle_{SW}\,, 
\end{equation}

\noindent once we identify the constants $\sqrt{c} $ and $ \Lambda $ that represent the QCD scale.

As we can see from eq. (\ref{Imag}), there is a non trivial compensation between the factors coming from the masses (\ref{nova2}) (\ref{nova3}) and the matrix elements (\ref{SmatrixComp})  in such a way that the leading order structure functions for soft and hard wall models are the same.
 
In the elastic limit, as we discussed in section {\bf 3} the form factor obtained from
the hadronic current in the soft wall model differs from the hard wall result. 
In refs. \cite{Grigoryan:2007my,Brodsky:2007hb} it was shown that the soft wall form factors are in better agreement with results obtained from experimental data.

We discuss in the appendix the problem of fermions in a dilaton background.
There we consider a hybrid model with hard and soft cut offs and find the same structure functions as in the hard wall model.
It would be interesting to calculate the structure functions in other holographic models for QCD such as \cite{Andreev:2006ct,Gursoy:2007cb,Gursoy:2007er}.

\bigskip

\noindent {\bf Acknowledgments:} The authors are partially supported by CLAF, CNPq and FAPERJ.

\appendix\section{A hybrid model for fermions }

The model that we considered in this article can not be used to fermions.
This happens because the dilaton background does not change the form of the fermionic free equation of motion
and does not lead to normalizable fermionic solutions.
Nevertheless, the presence of the dilaton background changes the solution for the virtual photon. So it is interesting to see the effect of this background on the structure functions in the fermionic case. 
For this purpose we consider here a different ten dimensional model:
an $AdS_5$  slice (hard cut off) times a five dimensional compact space with a dilaton background (soft cut off). Note that the virtual photon field is a non-normalizable solution so that it is not affected by the hard cut off. Then we can use the virtual photon solution obtained in section {\bf 3}.    
 
For a fermionic field in the dilaton background the $AdS_5$ sector of the action is proportional to
\begin{equation}
\int d^5 x \sqrt{ g_5}\,\, e^{-\varphi} \, 
 \bar\psi \Big( \frac{D}{2} - \frac{\stackrel{ \leftarrow}{D}}{2} \,-\,m_5 \,\Big) \,  \psi\, 
\end{equation}

\noindent where $g_5$ is the determinant of the $AdS_5$ metric and $\varphi = cz^2\,$.
The operators $D$ and $\stackrel{ \leftarrow}{D}$ are defined by
\begin{eqnarray}
D &\equiv&  \frac{z}{R}\, {\hat\gamma}^{ m}\, \partial_{ m} \,-\, \frac{2}{R} \, 
{\hat \gamma}^z
\nonumber\\
\stackrel{ \leftarrow}{D} &\equiv& \stackrel{ \leftarrow}{\partial}_{ m} \,
 \frac{z}{R}\, {\hat\gamma}^{ m}\,  \,-\,  \frac{2}{R}\, {\hat \gamma}^z
 \end{eqnarray}

\noindent where $\hat\gamma^m $ with $m=z,\mu\,$ are defined on the five dimensional tangent space with metric ${\rm diag} (1,-1,1,1,1)\,$, while $\gamma^m $ are defined in the curved AdS space. These matrices are related by $\hat\gamma^m \,=\, \frac{R}{z} \,\gamma^m $.

The equation of motion is
\begin{equation}
\Big( D  - m_5  - \frac{z}{2R} \partial_z \varphi \, \hat\gamma^z \Big) \psi\,=\,0\,\,.
\end{equation}

\noindent The five dimensional solution with a four dimensional plane wave factor with momentum $p$ and spin $\sigma$ is
\begin{equation}
\psi \,=\, C \, e^{ip \cdot y}\, e^{\varphi/2}\, z^{5/2}\, \Big[ \, J_{m_5R -1/2} (\sqrt{-p^2} z) P_+ \,
+ J_{m_5R + 1/2} ( \sqrt{-p^2 }z ) \,P_- \, \Big] \, u_\sigma
\label{fermionsol}
\end{equation}

\noindent where the Dirac spinor $u_\sigma$ satisfies 
$\,\,\slash\!\!\!{p}\, u_\sigma \,=\, \sqrt{-p^2}\, u_\sigma\,\,$, 
$\,\,P_{\pm} \,\equiv \, \frac{1}{2}( 1 \pm \hat\gamma^z)\,\,$ 
and $C$ is a normalization constant. The form of the fermionic solution is analogous to the hard wall solution. The dilaton shows up just as a multiplicative factor 
which cancels in the action and in the normalization condition. 
That means: the dilaton background alone does not work as an infrared cut off for the fermionic field. 
The normalization condition for the fermions is guaranteed by the hard cut off
$ z = z_{max}=1/\Lambda$ in the space.  

Considering ten dimensional fermionic fields of the form $\lambda\,=\, \psi(z,y) \,\otimes\, \eta (\Omega )\,$
with angular normalization: 
\begin{equation}
\int d^5 \Omega \sqrt{g_W} \bar\eta (\Omega ) \eta (\Omega ) \,=\, 1 \,,
\end{equation}

\noindent we find the normalization condition for $\psi$
\begin{equation}
R^{10}\int_0^{1/\Lambda}  \frac{dz}{z^5}\, e^{-\varphi}\, \bar\psi\, \gamma^1 \psi \,=\, 1 \,.
\end{equation}
 
\noindent This implies that the normalization constant for the fermionic solution (\ref{fermionsol}) reads 
\begin{equation}
C \,=\, \tilde C \, \frac{\Lambda^{1/2}}{R^{9/2}} (-p^2)^{1/4}\,,
\end{equation}

\noindent where $\tilde C$ is a dimensionless constant. 
The mass spectrum of the fermion field is discrete due to the boundary conditions at the hard cut off. They are not affected by the presence of the dilaton. This spectrum is determined from the zeros of the Bessel functions. Asymptotically, this implies 
\begin{equation}
 \sqrt{-p^2}=m_n=n\pi\Lambda\,.
\end{equation}

For the initial fermionic state with momentum $p=P$ and spin $\sigma$ and  mass $m_i\approx \pi\Lambda$ we can approximate in the interaction region $z\leq z_{int}\sim 1/q$
\begin{equation}
\psi_i \,=\,   \, \frac{\tilde C_i}{\Lambda^{3/2}R^{9/2}}  \,e^{i P \cdot y}\, e^{\varphi/2}\, (\Lambda z)^{\tau +1/2}\,  \, P_+ \, \, u_{i\sigma}\,\,,
\label{fermioni}
\end{equation}

\noindent where $\,\,\tau=\Delta-1/2\,\,$ and 
\begin{equation}
 \Delta = m_5\,R + 2 \,\,
\end{equation}

\noindent is the conformal dimension of the boundary operator.

The final fermionic state with momentum $p=P_X$ and spin $\sigma\prime$ can be written as
\begin{equation}
\psi_X \,=\, \tilde C_X \left(\frac{\Lambda}{R^9}\right)^{1/2} s^{1/4} \,e^{iP_X \cdot y}\, e^{\varphi/2}\, z^{5/2}\, \Big[ \, J_{\tau-2} (s^{1/2} z) P_+ \,
+ J_{\tau -1} ( s^{1/2} z ) \,P_- \, \Big] \, u_{X\sigma\prime}\,.
\label{fermionX}
\end{equation}

The fermion-photon interaction in the supergravity approximation is given by
\begin{eqnarray}
S_{int} = i\, {\cal Q}\, \int d^{10}x \sqrt{-g} e^{-\varphi} 
\, A_m \bar\lambda_X \, \gamma^m \, \lambda_i \,.
\end{eqnarray}

\noindent For simplicity we choose a polarization $\eta$ where $A_z=0$. Then, the interaction term reduces to 
\begin{eqnarray}
S_{int} &=& i\, {\cal Q}\, \int d^{4}y  \, dz \, d^5\Omega  \sqrt{-g} e^{-\varphi} 
\, A_\mu \bar\lambda_X \, \gamma^\mu \, \lambda_i \, \nonumber\\
&=& i\, {\cal Q}\,(2\pi)^4 \delta^4(P+q-P_X) \tilde C_i \tilde C_X \Lambda^{\tau-1/2}
s^{1/4} \, c \, \eta_\mu \, \bar{u}_{X\sigma\prime}\, \hat\gamma^\mu P_+ u_{i\sigma}\,\,
{\cal I}\,\,,
\end{eqnarray}

\noindent where
\begin{eqnarray}
{\cal I} \, = \, \Gamma\left( 1+\frac{q^2}{4c} \right) \, 
\int_0^{1/\Lambda} \, dz \, z^{\tau+1}
\,\,{\cal U}  ( 1 + \frac{q^2}{4c} ; 2 ; cz^2 ) \,\, J_{\tau-2} (s^{1/2} z)\,\,.
\end{eqnarray}

The integral in $\cal I$ can be rewritten defining $\,\zeta=(qz)^2/4\,$ and 
$\,a= 1 + q^2/4c \,$
\begin{eqnarray}
{\cal I} \, 
= \,\frac 12 \left(\frac{2}{q}\right)^{\tau+2} 
\int_0^{c(a-1)/\Lambda^2} \, d\zeta \, \zeta^{\tau/2}\, \Gamma\left( a \right) 
\,\,{\cal U}  \left( a ; 2 ; \frac{\zeta}{a-1} \right) \,\, 
J_{\tau-2} \left( 2 \zeta^{1/2} \sqrt{\frac 1x -1} \right)\,\,.\nonumber\\
\end{eqnarray}

The fermionic structure functions can be obtained from this integral.
We will just consider the leading order contribution in the DIS limit.
The integrand of the above expression is negligible for any $\,\,\zeta\,\geq\, \zeta_0 \,\,$ such that  $1 << \zeta_0 << c(a-1)/\Lambda^2 = q^2/4\Lambda^2$. Then, in the DIS limit ($q^2\to \infty$ with $x$ fixed) $a\to\infty$ and $\zeta$ is bounded so we can approximate the confluent hypergeometric function  as  
\begin{eqnarray}
\Gamma\left( a \right) \,\,{\cal U}  \left( a ; 2 ; \frac{\zeta}{a-1} \right) \,\, 
\approx \, 2 \, (a-1)\, \zeta^{-1/2} K_{1} \left( 2 \, \zeta^{1/2}  \right)\,\,, 
\end{eqnarray}

\noindent and we find that 
\begin{eqnarray}
{\cal I} \, 
\approx  \, \frac {\Gamma(\tau)}{2c} \,  \, \left(\frac{2x}{q}\right)^{\tau} 
\, \left(\frac 1x -1\right)^{\frac \tau{2}-1}\,\,.
\end{eqnarray}

In a similar way to the scalar case, we can then extract the matrix elements of the current from the interaction term as
\begin{eqnarray}
\eta_\mu \langle P_X, X, \sigma^\prime \vert J^\mu ( 0 )  \vert P,{\cal Q}, \sigma \rangle 
&=& i\, {\cal Q}\, \tilde C_i \tilde C_X \Lambda^{\tau-1/2}
s^{1/4} \, c \, 
 \, \eta_\mu \, \bar{u}_{X\sigma\prime}\, \hat\gamma^\mu P_+ u_{i\sigma}\,\,
 \nonumber\\ 
&& \times \frac {\Gamma(\tau)}{2c} \,  \, \left(\frac{2x}{q}\right)^{\tau} 
\, \left(\frac 1x -1\right)^{\frac \tau{2}-1}\,\,
\end{eqnarray}

In order to obtain the imaginary part of the forward Compton amplitude we have to sum  over radial excitations and final spins and average over initial spins. We find 
\begin{eqnarray}
\eta_\mu \, \eta_\nu \,{\rm Im}\, T^{\mu\nu}  
&=&  {\cal Q}^2 C^\prime \Lambda^{2\tau-2} x^{\tau + 2} (1-x)^{\tau-2} q^{-2\tau} 
 \Big[ \, (P\cdot \eta)^2 - \frac 12 P\cdot q \, \eta^2 \Big]\,\,,
\end{eqnarray}

\noindent where $\, C^\prime \,=\, 2^{2\tau-1}\,[\tilde C_i\, \tilde C_X \, \Gamma(\tau)]^2\, $. Then, we obtain the leading order structure functions in the fermionic case
\begin{eqnarray}
F_2\,=\,2\,F_1 
&=& \pi \, {\cal Q}^2 \, C^\prime \, \left(\frac{\Lambda^2}{q^2}\right)^{\tau-1} \,
 x^{\tau + 1} \, (1-x)^{\tau-2} \,\,,
\end{eqnarray}

\noindent in agreement with the hard cut off calculation presented in ref. 
\cite{Polchinski:2002jw}. Note that in spite of the photon dependence on the dilaton scale $\sqrt{c}$, the above result depends only on the hard cut off scale $\Lambda$ of this hybrid model.


\begin{thebibliography}{ABC}


\bibitem{Maldacena:1997re}
  J.~M.~Maldacena,
  %``The large N limit of superconformal field theories and supergravity,''
  Adv.\ Theor.\ Math.\ Phys.\  {\bf 2}, 231 (1998)
  [Int.\ J.\ Theor.\ Phys.\  {\bf 38}, 1113 (1999)].
 [arXiv:hep-th/9711200].
  %%CITATION = HEP-TH 9711200;%%

%\cite{Gubser:1998bc}
\bibitem{Gubser:1998bc}
  S.~S.~Gubser, I.~R.~Klebanov and A.~M.~Polyakov,
  %``Gauge theory correlators from non-critical string theory,''
  Phys.\ Lett.\  B {\bf 428}, 105 (1998).
  [arXiv:hep-th/9802109].
  %%CITATION = PHLTA,B428,105;%%

%\cite{Witten:1998qj}
\bibitem{Witten:1998qj}
  E.~Witten,
  %``Anti-de Sitter space and holography,''
  Adv.\ Theor.\ Math.\ Phys.\  {\bf 2}, 253 (1998).
  [arXiv:hep-th/9802150].
  %%CITATION = 00203,2,253;%%

%\cite{Klebanov:2000hb}
\bibitem{Klebanov:2000hb}
  I.~R.~Klebanov and M.~J.~Strassler,
  %``Supergravity and a confining gauge theory: Duality cascades and
  %chiSB-resolution of naked singularities,''
  JHEP {\bf 0008}, 052 (2000)
  [arXiv:hep-th/0007191].
  %%CITATION = JHEPA,0008,052;%% 


%\cite{Maldacena:2000yy}
\bibitem{Maldacena:2000yy}
  J.~M.~Maldacena and C.~Nunez,
  %``Towards the large N limit of pure N = 1 super Yang Mills,''
  Phys.\ Rev.\ Lett.\  {\bf 86}, 588 (2001)
  [arXiv:hep-th/0008001].
  %%CITATION = PRLTA,86,588;%%


%\cite{Polchinski:2001tt}
\bibitem{Polchinski:2001tt}
  J.~Polchinski and M.~J.~Strassler,
  %``Hard scattering and gauge/string duality,''
  Phys.\ Rev.\ Lett.\  {\bf 88}, 031601 (2002)
  [arXiv:hep-th/0109174].
  %%CITATION = PRLTA,88,031601;%%


%\cite{Polchinski:2002jw}
\bibitem{Polchinski:2002jw}
  J.~Polchinski and M.~J.~Strassler,
  %``Deep inelastic scattering and gauge/string duality,''
  JHEP {\bf 0305}, 012 (2003)
  [arXiv:hep-th/0209211].
  %%CITATION = JHEPA,0305,012;%%


\bibitem{QCD1} V. A. Matveev, R.M. Muradian and A. N. Tavkhelidze,
 Lett. Nuovo Cim. {\bf 7} (1973) 719.

\bibitem{BRO} S. J. Brodsky and G. R. Farrar, Phys. Rev. Lett 31 (1973) 1153;
Phys. Rev. {\bf D11} (1975) 1309.


%\cite{Giddings:2002cd}
\bibitem{Giddings:2002cd}
  S.~B.~Giddings,
  %``High energy QCD scattering, the shape of gravity on an IR brane, and  the
  %Froissart bound,''
  Phys.\ Rev.\  D {\bf 67}, 126001 (2003)
  [arXiv:hep-th/0203004].
  %%CITATION = PHRVA,D67,126001;%%

%\cite{BoschiFilho:2002zs}
\bibitem{BoschiFilho:2002zs}
  H.~Boschi-Filho and N.~R.~F.~Braga,
  %``QCD / string holographic mapping and high energy scattering amplitudes,''
  Phys.\ Lett.\  B {\bf 560}, 232 (2003)
  [arXiv:hep-th/0207071].
  %%CITATION = PHLTA,B560,232;%%

%\cite{Brower:2002er}
\bibitem{Brower:2002er}
  R.~C.~Brower and C.~I.~Tan,
  %``Hard scattering in the M-theory dual for the QCD string,''
  Nucl.\ Phys.\  B {\bf 662}, 393 (2003)
  [arXiv:hep-th/0207144].
  %%CITATION = NUPHA,B662,393;%%

%\cite{Andreev:2002aw}
\bibitem{Andreev:2002aw}
  O.~Andreev,
  %``Scaling laws in hadronic processes and string theory,''
  Phys.\ Rev.\  D {\bf 67}, 046001 (2003)
  [arXiv:hep-th/0209256].
  %%CITATION = PHRVA,D67,046001;%%
 
%\cite{Boschi-Filho:2002vd}
\bibitem{Boschi-Filho:2002vd}
  H.~Boschi-Filho and N.~R.~F.~Braga,
   JHEP {\bf 0305}, 009 (2003) [arXiv:hep-th/0212207];
  Eur.\ Phys.\ J.\  C {\bf 32}, 529 (2004)[arXiv:hep-th/0209080].

 %%CITATION = EPHJA,C32,529;%%

%\cite{deTeramond:2005su}
\bibitem{deTeramond:2005su}
  G.~F.~de Teramond and S.~J.~Brodsky,
  %``The hadronic spectrum of a holographic dual of QCD,''
  Phys.\ Rev.\ Lett.\  {\bf 94}, 201601 (2005)
  [arXiv:hep-th/0501022].
  %%CITATION = PRLTA,94,201601;%%

%\cite{Erlich:2005qh}
\bibitem{Erlich:2005qh}
  J.~Erlich, E.~Katz, D.~T.~Son and M.~A.~Stephanov,
%``QCD and a holographic model of hadrons,''
Phys. Rev. Lett. {\bf 95} (2005) 261602
[arXiv:hep-ph/0501128].

%\cite{DaRold:2005zs}
\bibitem{DaRold:2005zs}
  L.~Da Rold and A.~Pomarol,
%``Chiral symmetry breaking from five dimensional spaces,''
 Nucl.\ Phys.\  {\bf B 721} (2005) 79
[arXiv:hep-ph/0501218].

%\cite{Karch:2006pv}
\bibitem{Karch:2006pv}
  A.~Karch, E.~Katz, D.~T.~Son and M.~A.~Stephanov,
  %``Linear confinement and AdS/QCD,''
  Phys.\ Rev.\  D {\bf 74}, 015005 (2006)
  [arXiv:hep-ph/0602229].
  %%CITATION = PHRVA,D74,015005;%%
 
%\cite{Colangelo:2007pt}
\bibitem{Colangelo:2007pt}
  P.~Colangelo, F.~De Fazio, F.~Jugeau and S.~Nicotri,
  %``On the light glueball spectrum in a holographic description of QCD,''
  Phys.\ Lett.\  B {\bf 652}, 73 (2007)
  [arXiv:hep-ph/0703316].
  %%CITATION = PHLTA,B652,73;%%

%\cite{Forkel:2007cm}
\bibitem{Forkel:2007cm}
  H.~Forkel, M.~Beyer and T.~Frederico,
  %``Linear square-mass trajectories of radially and orbitally excited hadrons
  %in holographic QCD,''
  JHEP {\bf 0707}, 077 (2007)
 [arXiv:0705.1857 [hep-ph].
  %%CITATION = JHEPA,0707,077;%%

%\cite{Herzog:2006ra}
\bibitem{Herzog:2006ra}
  C.~P.~Herzog,
  %``A holographic prediction of the deconfinement temperature,''
  Phys.\ Rev.\ Lett.\  {\bf 98}, 091601 (2007)
  [arXiv:hep-th/0608151].
  %%CITATION = PRLTA,98,091601;%%

%\cite{BallonBayona:2007vp}
\bibitem{BallonBayona:2007vp}
  C.~A.~Ballon Bayona, H.~Boschi-Filho, N.~R.~F.~Braga and L.~A.~Pando Zayas,
  %``On a holographic model for confinement / deconfinement,''
Phys. \ Rev. D {\bf 77}, 046002 (2008)
  [arXiv:0705.1529 [hep-th]].
  %%CITATION = ARXIV:0705.1529;%%

%\cite{Manohar:1992tz}
\bibitem{Manohar:1992tz} For a review see:
  A.~V.~Manohar,
  %``An introduction to spin dependent deep inelastic scattering,''
  arXiv:hep-ph/9204208.
  %%CITATION = HEP-PH/9204208;%%

%\cite{Hatta:2007he}
\bibitem{Hatta:2007he}
  Y.~Hatta, E.~Iancu and A.~H.~Mueller,
  %``Deep inelastic scattering at strong coupling from gauge/string duality :
  %the saturation line,''
  JHEP {\bf 0801}, 026 (2008)
  [arXiv:0710.2148 [hep-th]].
  %%CITATION = JHEPA,0801,026;%%

%\cite{Grigoryan:2007vg}
\bibitem{Grigoryan:2007vg}
  H.~R.~Grigoryan and A.~V.~Radyushkin,
  %``Form Factors and Wave Functions of Vector Mesons in Holographic QCD,''
  Phys.\ Lett.\  B {\bf 650}, 421 (2007)
  [arXiv:hep-ph/0703069].
  %%CITATION = PHLTA,B650,421;%%

%\cite{Grigoryan:2007my}
\bibitem{Grigoryan:2007my}
  H.~R.~Grigoryan and A.~V.~Radyushkin,
  %``Structure of Vector Mesons in Holographic Model with Linear Confinement,''
  Phys.\ Rev.\  D {\bf 76}, 095007 (2007)
  [arXiv:0706.1543 [hep-ph]].
  %%CITATION = PHRVA,D76,095007;%%

%\cite{Brodsky:2007hb}
\bibitem{Brodsky:2007hb}
  S.~J.~Brodsky and G.~F.~de Teramond,
  %``Light-Front Dynamics and AdS/QCD: The Pion Form Factor in the Space- and
  %Time-Like Regions,''
  arXiv:0707.3859 [hep-ph].
  %%CITATION = ARXIV:0707.3859;%%

%\cite{Hatta:2007cs}
\bibitem{Hatta:2007cs}
  Y.~Hatta, E.~Iancu and A.~H.~Mueller,
  %``Deep inelastic scattering off a N=4 SYM plasma at strong coupling,''
  JHEP {\bf 0801}, 063 (2008)
  [arXiv:0710.5297 [hep-th]].


%\cite{Andreev:2006ct}
\bibitem{Andreev:2006ct}
  O.~Andreev and V.~I.~Zakharov,
  %``Heavy-quark potentials and AdS/QCD,''
  Phys.\ Rev.\  D {\bf 74}, 025023 (2006)
  [arXiv:hep-ph/0604204].
  %%CITATION = PHRVA,D74,025023;%%

%\cite{Gursoy:2007cb}
\bibitem{Gursoy:2007cb}
  U.~Gursoy and E.~Kiritsis,
  %``Exploring improved holographic theories for QCD: Part I,''
  arXiv:0707.1324 [hep-th].
  %%CITATION = ARXIV:0707.1324;%%

%\cite{Gursoy:2007er}
\bibitem{Gursoy:2007er}
  U.~Gursoy, E.~Kiritsis and F.~Nitti,
  %``Exploring improved holographic theories for QCD: Part II,''
  arXiv:0707.1349 [hep-th].
  %%CITATION = ARXIV:0707.1349;%%





 


%%%%%%%%%%%%%%%%%%%%%%%%%%%%%%%%%%%%%%%%%%%%

 

\end{thebibliography}
 \end{document}